\documentclass[preprint,aps,showpacs,nofootinbib,preprintnumbers,amsmath,amssymb]{revtex4-1}
\usepackage{epsfig}
\usepackage{subfigure}
\usepackage{multirow}
\usepackage{dcolumn}
\usepackage{bm}
\usepackage[usenames ,dvipsnames]{xcolor}
\usepackage{slashed}
\usepackage{float}
\usepackage{subfigure}
\usepackage{graphicx,color}
\usepackage{amsmath}
\usepackage{slashed}
\usepackage{nonfloat}
\begin{document}
\title{ Non-leptonic two-body decays  of $\Lambda_b^0$ in light-front  quark model}

\author{C.Q. Geng$^{1,2,3,4}$, Chia-Wei Liu$^{3}$ and Tien-Hsueh Tsai$^{3}$}
\affiliation{
$^{1}$School of Fundamental Physics and Mathematical Sciences, Hangzhou Institute for Advanced Study, UCAS, Hangzhou 310024, China \\
$^{2}$International Centre for Theoretical Physics Asia-Pacific, Beijing/Hangzhou, China \\
$^{3}$Department of Physics, National Tsing Hua University, Hsinchu 300, Taiwan\\
$^{4}$Physics Division, National Center for Theoretical Sciences, Hsinchu 300, Taiwan
}\date{\today}

\begin{abstract}
We study  the   non-leptonic two-body weak decays  of $\Lambda_b^0 \to p M$ with $ M=(\pi^-,K^-)$ and $(\rho^-,K^{*-})$  in
the  light-front  quark model   under the generalized factorization ansatz. 
 By considering the Fermi statistic between quarks and determining spin-flavor structures  in baryons,  we  calculate the branching ratios (${\cal B}$s) 
 and CP-violating rate asymmetries ($\mathcal{A}_{CP}$s) in the decays.
Explicitly, we find that 
${\cal B}( \Lambda_b^0 \to p \pi^- ,pK^-)=(4.18\pm0.15\pm0.30, 5.76\pm0.88\pm0.23)\times10^{-6}$ and
${\mathcal{A}_{CP}}( \Lambda_b^0 \to p \pi^- ,\,pK^-)=(-3.60\pm0.14\pm0.14, 6.36\pm0.21\pm0.18)\%$  
in comparison  with the data of ${\cal B}( \Lambda_b^0 \to p \pi^- ,pK^-)=(4.5\pm0.8, 5.4\pm1.0)\times10^{-6}$ and
${\mathcal{A}_{CP}}( \Lambda_b^0 \to p \pi^- ,pK^-)=(-2.5\pm 2.9, -2.5\pm2.2)\%$
 given by the Particle Data Group, respectively.
We also predict that 
${\cal B}(  \Lambda_b^0 \to p \rho^-,pK^{*-} )=(12.13\pm3.27\pm0.91, 2.58\pm0.87\pm0.13)\times 10^{-6}$ and
${\mathcal{A}_{CP}}( \Lambda_b^0 \to p \rho^-,pK^{*-} )=(-3.32\pm0.00\pm0.14,19.25\pm0.00\pm0.80)\%$,  which could be observed by the  experiments at  LHCb.
\end{abstract}
\maketitle

\section{introduction}
It is known that $b$-physics has been providing us with  a nature platform to observe 
CP-violating phenomena, test the heavy quark effective theory, and explore  physics beyond the Standard Model (SM).  
Among the various processes,  the weak decays of  $\Lambda_{b}^0$ as a complementary of  the $B$-meson ones
 give us an opportunity to verify the QCD factorization  hypothesis. 
  In the recent years, several interesting $\Lambda_{b}^0$  decay processes have been measured by the LHCb Collaboration, 
  such as the two-body non-leptonic mode of $\Lambda_{b}^0\to \Lambda \phi$~\cite{Aaij:2016zhm} and  the radiative decay of
   $\Lambda_{b}^0\to \Lambda \gamma$~\cite{Aaij:2019hhx}.
   In addition,  the direct CP violating rate asymmetries in  $\Lambda_{b}^0\to p \pi^-$ and  $\Lambda_{b}^0\to p K^-$
   have also been searched with the most recent data of $(-2.5\pm2.9)\%$ and $(-2.5\pm2.2)\%$~\cite{Aaij:2018tlk,pdg}, respectively.
  One  expects that  LHCb will accumulate more and more high quality data after the upgrade and lead b-physics into a precision era.  
  There is no doubt that comprehensive studies in the   $\Lambda_{b}^0$ processes are necessary. Particularly,  only theoretical
  estimations could not be able to understand the future LHCb experimental measurements.   
  To describe the exclusive $\Lambda_{b}^0$ decays, there are lots of different QCD approaches, such as the generalized   factorization approach (GFA)~\cite{HG,Hsiao:2017tif},  
 perturbative QCD (pQCD)  method~\cite{pQCD},   light-front quark model (LFQM)~\cite{Zhu:2018jet,Zhao:2018zcb},
 MIT bag model (MBM)~\cite{Geng:2020ofy}, and light cone sum rule (LCSR)~\cite{Khodjamirian:2011jp}. 
  
  In this work, we concentrate on the non-leptonic two body decays of $\Lambda_{b}^0\to p M$  with $M$ representing a pseudoscalar (P) or vector (V) meson in the final states. We use the effective Hamiltonian including the QCD-penguin and electroweak-penguin operators and the effective Wilson coefficients are evaluated at the renormalization scale $\mu=2.5 \text{ GeV}$ in the NLL precision~\cite{Buras:1991jm,Ali:1998eb}. To obtain the decay amplitudes, we follow GFA to split the matrix elements into two pieces, resulting in that the only relevant quantities for the decay amplitudes are meson decay constants and baryon transition form factors. The hardest part to get the decay amplitude is to calculate the baryonic transition form factors in $\Lambda_b^0\to p$
   because of  the complicated baryon structures and the non-perturbative nature of QCD at the low energy scale.   
   To extract the form factors, we use LFQM, which has been widely used in the $B$-meson~\cite{Geng:1997ws,Geng:2001de,Geng:2003su,Geng:2016pyr} and heavy to heavy baryonic transition
  systems~\cite{Chua:2018lfa,Chua:2019yqh}. The greatest advantage of LFQM is that we can deal the baryon states consistently 
  with different momenta because of the boost invariance property in the light front dynamics. As a trade off, we are only allowed to evaluate the form factors  
  in the space-like region to avoid the zero-modes or so called Z-graphs, which are hard to be calculated in LFQM. 
  We analyze the branching ratios of 
  the pseudoscalar modes for $\Lambda_{b}^0\to (p \pi^-, pK^-)$ and their corresponding CP-violating
  asymmetries, and compare them with the current experimental data. 
  We also predict  the  vector decay modes of $\Lambda_{b}^0\to (p \rho^-,pK^{*-})$. In particular, we would like to check if 
   the sizable CP-violating rate asymmetry in $\Lambda_{b}^0 \to p K^{*-}$, predicted to be as large as $20\%$ in GFA~\cite{HG,Hsiao:2017tif},
   can be confirmed in LFQM.
  
This paper is organized as follows. In Sec.~II, we present our formalism, which contains the defective Hamiltonians, decay widths and asymmetries, vertex
functions of the baryons and baryonic transition form factors in LFQM.
  We show our numerical results of the form factors, branching ratios and CP asymmetries and compare our results  with those in the literature in Sec. III.
In Sec.~IV,  we give our  conclusions.

\section{Formalism}
\subsection{Effective Hamiltonians}
To study 
 the exclusive two-body non-leptonic processes of $\Lambda_b^0\to pM$ with $M$ being the pesudoscalar $P=(\pi^-,K^-)$ and vector $V=(\rho^-,K^{*-})$
 mesons,
 we start with  the effective Hamiltonians of $b\to qu\bar{u}$ ($q=d,s$) at quark level, given by
\begin{eqnarray}
	\mathcal{H}_{e f f}=\frac{G_{F}}{\sqrt{2}}\left[V_{u b} V_{u q}^{*}\left(C_{1} O_{1}+C_{2} O_{2}\right)-V_{t b} V_{t q}^{*}\sum_{i=3}^{10} C_{i} O_{i}\right] \,,
\end{eqnarray}
where $G_F$ is the  Fermi constant, $C_i$  stand for  the  Wilson coefficients evaluated at the renormalization scale $\mu$, $V_{q_1q_2}$ represent the CKM quark mixing matrix elements, and $O_{1-10}$ are the operators,  given as 
\begin{eqnarray}
	O_1&=&(\bar{q}u)_{V-A}(\bar{u}b)_{V-A},\qquad \qquad\qquad	O_2=(\bar{q}_\beta u_\alpha)_{V-A}(\bar{u}_\alpha b_\beta)_{V-A} \nonumber\\
	O_3&=&(\bar{q}b)_{V-A}\sum_{Q}(\bar{Q}Q)_{V-A},\qquad \qquad	O_4=(\bar{q}_\beta b_\alpha)_{V-A}\sum_{Q}(\bar{Q}_\alpha Q_\beta)_{V-A} \nonumber\\
	O_5&=&(\bar{q}b)_{V-A}\sum_{Q}(\bar{Q}Q)_{V+A},\qquad \qquad	O_6=(\bar{q}_\beta b_\alpha)_{V-A}\sum_{Q}(\bar{Q}_\alpha Q_\beta)_{V+A} \nonumber\\
	O_7&=&\frac{3}{2}(\bar{q}b)_{V-A}\sum_{Q}e_{Q}(\bar{Q}Q)_{V+A},\qquad 	O_8=\frac{3}{2}(\bar{q}_\beta b_\alpha)_{V-A}\sum_{Q}e_{Q}(\bar{Q}_\alpha Q_\beta)_{V+A} \nonumber\\
O_9&=&\frac{3}{2}(\bar{q}b)_{V-A}\sum_{Q}e_{Q}(\bar{Q}Q)_{V-A},\qquad 	O_{10}=\frac{3}{2}(\bar{q}_\beta b_\alpha)_{V-A}\sum_{Q}e_{Q}(\bar{Q}_\alpha Q_\beta)_{V-A} \,.
\end{eqnarray}
Here, $O_{1,2}$, $O_{3-6}$ and  $O_{7-10}$ correspond to the   tree, QCD  and   electroweak-penguin loop operators, 
respectively, while $Q=u,d,s,c,b$ for  $\mu=O(m_b)$.
To calculate the decays of $\Lambda_b^0\to pM$, we need to find the matrix elements for the operators, given by
\begin{eqnarray}
\label{MatrixE}
	\langle pM|\mathcal{H}_{eff}|\Lambda_b^0 \rangle=\sum_{i,j}c_{i}^{eff}(\mu)\langle pM |O_j|\Lambda_b^0\rangle\,,
\end{eqnarray}
where the explicit expressions for the effective Wilson coefficient $c_i^{eff}(\mu)$ can be found in Refs.~\cite{Ali:1998eb,Buras:1991jm}.
Note that the physical matrix elements on the left-handed side of  Eq.~(\ref{MatrixE}) should be independent of
the renormalization scheme and scale.

Based on  GFA,
each element  can be written as
\begin{eqnarray}\label{fact}
	\langle p M|O_i|\Lambda_b^0\rangle=\langle M|(\bar{q}u)|0\rangle\langle p| (\bar{u} b)|\Lambda_b^0\rangle\,,
\end{eqnarray}
 where the explicit Dirac and color indices are suppressed.
As a result,  the decay amplitudes are govern by the mesonic and baryonic transitions separately.
For the former, we can use the definitions, given by
\begin{eqnarray}
\langle 0|A^{\mu}|P\rangle=if_P p_P^{\mu},\qquad 	\langle 0|A^{\mu}|V\rangle=f_Vm_V \epsilon^{\mu}\,,
\end{eqnarray}
where  $f_M$ ($M=P,V$) are the meson decay constants,
$m_M$ correspond to the masses of $M$, and $p^\mu(\epsilon^{\mu})$ is the momentum (polarization) vector for $P(V)$.
The latter can be related to the baryonic transition form factors, defined by
\begin{eqnarray}\label{form}
\langle p|V_{\mu}-A_{\mu}|\Lambda_b\rangle&=&\bar{u}_{p}(p_f)\left(f_1(k^2)\gamma_{\mu}-i\sigma_{\mu\nu}\frac{k^\nu}{m_{\Lambda_b}}f_2(k^2)+\frac{k_\mu}{m_{\Lambda_b}}f_3(k^2)\right)u_{\Lambda_b}(p_i)\nonumber\\
&-&\bar{u}_{p}(p_f)\left(g_1(k^2)\gamma_{\mu}-i\sigma_{\mu\nu}\frac{k^\nu}{m_{\Lambda_b}}g_2(k^2)+\frac{k_\mu}{m_{\Lambda_b}}g_3(k^2)\right)\gamma_5u_{\Lambda_b}(p_i)\,,
\end{eqnarray}
where $k^{\mu}=p_i^{\mu}-p_f^{\mu}$ and $k^2=m^2_{P(V)}$. The matrix elements for $O_{5-8}$, which have the $(V-A)(V+A)$ structure,
 can be calculated by the employment of the Dirac equation after applying the Fierz transformation and factorization, given by
\begin{eqnarray}
	\langle p M|(V-A)(V+A)|\Lambda_b^0\rangle&=&-2\langle M|(S+P)|0\rangle\langle p|(S-P)|\Lambda_b^0\rangle\nonumber\\
	&=&-\left[R_1\langle p|V_\mu|\Lambda_b^0\rangle+R_2\langle p|A_\mu|\Lambda_b^0\rangle\right]\langle M|(V-A)^{\mu}|0\rangle
\end{eqnarray}
with
\begin{eqnarray}
	R_1=\frac{2m_M^2}{(m_b-m_u)(m_q+m_u)} \quad 	R_2=\frac{2m_M^2}{(m_b+m_u)(m_q+m_u)}\,,
	\label{chiral enhancement}
\end{eqnarray}
where the quark masses are the current quark ones. 
Consequently, the amplitude for  $\Lambda_b^0\to p P$ is written as
\begin{eqnarray}\label{2bodyMexp}
	\mathcal{A}( \Lambda_b^0\to p P)&=&i\frac{G_F}{\sqrt{2}}f_{P}\bar{u}_{p}(p_f)\left\{ [V_{ub}V_{uq}^*a_1-V_{tb}V_{tq}^*(a_4+a_{10})][f_1(k^2)(m_{\Lambda_b}-m_p)\right.\nonumber\\
	&&+g_1(k^2)(m_{\Lambda_b}+m_p)\gamma_5]-V_{tb}V_{tq}^*(a_6+a_8) [R_1f_1(k^2)(m_{\Lambda_b}-m_p)
	\nonumber\\
	&&\left.-R_2g_1(k^2)(m_{\Lambda_b}-m_p)\gamma_5]\right \}u_{\Lambda_b}(p_i)\,,
\end{eqnarray}
where $a_i=c^{eff}_i+c^{eff}_{i+1(-1)}/N_c^{eff}$ for $i=\text{odd(even)}$ with the effective color number $N_c^{eff}$ to parameterize 
the non-factorizable QCD effects of the octet-octet operators  in Eq.~(\ref{fact}). 
Here, $f_2$ and $g_2$ have no contributions to the amplitude  due to  the anti-symmetric structure of $\sigma_{\mu\nu}$, while
 the terms associated with $f_3$ and $g_3$ are suppressed by the factor of $m_P^2/m_{\Lambda_b}^2$.
Similarly, the amplitude for  $\Lambda_b^0\to p V$ is given by
\begin{eqnarray}\label{2bodyVexp}
	\mathcal{A}( \Lambda_b^0\to p V)&=&i\frac{G_F}{\sqrt{2}}f_{V}m_V\epsilon^{*\mu}\bar{u}_{p}(p_f)\left\{\left[V_{ub}V_{uq}^*a_1-V_{tb}V_{tq}^*(a_4+a_{10})\right]\right.\nonumber\\
&&	
	\left[ \left(f_1(k^2)+f_2(k^2)\frac{m_{\Lambda_b}+m_p}{m_{\Lambda_b}}\right)\gamma_{\mu}
	-2f_2(k^2)\frac{(p_{f})_{\mu}}{m_{\Lambda_b}}\right.\nonumber\\
&&	
	\left.\left.-\left(g_1(k^2)-g_2(k^2)\frac{m_{\Lambda_b}-m_p}{m_{\Lambda_b}}\right)\gamma_{\mu}\gamma_5-2g_2(k^2)\frac{(p_f)_{\mu}}{m_{\Lambda_b}}\gamma_5\right]
	\right\}u_{\Lambda_b}(p_i)\,.
\end{eqnarray}

\subsection{Decay widths and CP asymmetries}
The decay widths of  $\Lambda_{b}^0\to p M$ ($M=P,V$) can be found from
Eqs.~(\ref{2bodyMexp}) and (\ref{2bodyVexp}), read as
\begin{eqnarray}\label{2bodyM}
\Gamma (\Lambda_{b}^0\to p P)&=&\frac{p_c}{16\pi}G_F^2f_P^2 \left\{  \frac{(m_{\Lambda_b}+m_p)^2-m_P^2}{m_{\Lambda_b}^2}\left|\Big[V_{ub}V_{uq}^*a_1 -V_{tb}V_{tq}^*[a_4+a_{10}\right. \right. \nonumber\\ 
 &&\left. +R_1(a_6+a_8)]\Big]\right|^2 f_1^2(m_P^2)(m_{\Lambda_b}-m_p)^2+\frac{(m_{\Lambda_{b}}-m_p)^2-m_P^2}{m_{\Lambda_{b}}^2} \nonumber  \\
&&\left.\left|\Big[V_{ub}V_{uq}^*a_1 -V_{tb}V_{tq}^*[a_4+a_{10}-R_2(a_6+a_8)]\Big] \right|^2g_1^2(m_P^2)(m_{\Lambda_b}+m_p)^2\right\}
\end{eqnarray}
and 
\begin{eqnarray}\label{2bodyV}
\Gamma (\Lambda_b ^0\to p V) &=&\frac{p_c}{8\pi}\frac{E_p+m_p}{m_{\Lambda_b}}G_F^2f_V^2m_V^2\Big|V_{ub}V_{uq}^*a_1 -V_{tb}V_{tq}^*(a_4+a_{10})\Big|^2\nonumber\\
&&\left\{2\left(g_1(m_V^2)-\frac{m_{\Lambda_b}-m_p}{m_{\Lambda_b}}g_2(m_V^2)\right)^2+2\frac{E_p-m_p}{E_p+m_p}\bigg( f_1(m_V^2)\bigg.\right.\nonumber\\
&&\left.+\frac{m_{\Lambda_b}+m_p}{m_{\Lambda_b}}f_2(m_V^2)\right)^2+\left(\frac{m_{\Lambda_b}-m_p}{m_V}g_1(m_V^2)-\frac{m_V}{m_{\Lambda_b}}g_2(m_V^2)\right)^2\nonumber\\ 
&&\left.+\frac{E_p-m_p}{E_p+m_p}\left(\frac{m_{\Lambda_b}+m_p}{m_V}f_1(m_V^2)+\frac{m_V}{m_{\Lambda_b}}f_2(m_V^2)\right)^2\right\}\,,\\
\nonumber
\end{eqnarray}
where $p_c$  is the  momentum in the center mass frame and $E_{p}$ is the energy of the proton.
The direct CP-violating rate asymmetry is defined by
\begin{eqnarray}
	{\cal A}_{CP}(\Lambda_b^0\to p M)\equiv\frac{\Gamma(\Lambda_b^0\to p M)-\Gamma(\bar{\Lambda}_b^0\to \bar{p} \bar{M})}{\Gamma(\Lambda_b^0\to p M)+\Gamma(\bar{\Lambda}_b^0\to \bar{p} \bar{M})}
	\label{CP}
\end{eqnarray}
where $\Gamma(\Lambda_b^0\to p M)$ and $\Gamma(\bar{\Lambda}_b^0\to \bar{p} \bar{M})$ are the decay widths of the particle and antiparticle, respectively.

\subsection{Vertex functions of baryons}
In LFQM, a baryon with its momentum $P$ and spin $S$ as well as the z-direction projection of spin $S_z$ are considered as a bound state of three constitute quarks.
As a result, the baryon state can be expressed by~\cite{Zhang:1994ti,Geng:2020fng,Schlumpf:1992ce,Cheng:2004cc,Ke:2007tg,Ke:2012wa,Ke:2019smy}
\begin{eqnarray}
\label{eq1}
	&|{\bf B}&,P,S,S_z\rangle=\int\{{d^3{\tilde{p}_1}}\}\{{d^3{\tilde{p}_2}}\}\{{d^3{\tilde{p}_3}}\}2(2\pi)^3\frac{1}{\sqrt{P^+}}\delta^3(\tilde{P}-\tilde{p}_1-\tilde{p}_2-\tilde{p}_3) \nonumber \\
	&\times& \sum_{\lambda_1,\lambda_2,\lambda_3}\Psi^{SS_z}(\tilde{p}_1,\tilde{p}_2,\tilde{p}_3,\lambda_1,\lambda_2,\lambda_3)C^{\alpha\beta\gamma}F_{abc}|q_{\alpha}^{a}(\tilde{p}_1,\lambda_1) q_{\beta}^{b}(\tilde{p}_2,\lambda_2)q_{\alpha}^{a}(\tilde{p}_3,\lambda_3) \rangle\,,
	\label{baryon}
\end{eqnarray}
where  $\Psi^{SS_z}(\tilde{p}_1,\tilde{p}_2,\tilde{p}_3,\lambda_1,\lambda_2,\lambda_3)$  is
 the vertex function, which can be formally solved from the Bethe-Salpeter equations by the Faddeev decomposition method,
$C^{\alpha\beta\gamma}$ ($F_{abc}$) is the color  (flavor) factor, $\lambda_i$  ($\tilde{p}_i$) with $i=1,2,3$  are the LF helicities 
(3-momenta) of the on-mass-shell constituent quarks, defined as
\begin{eqnarray}
\tilde{p}_{i}=(p_i^+,p_{i\perp})\,, ~p_{i\perp}=(p_i^1,p_i^2) \,,~ p_i^-=\frac{m_i^2+p_{i\perp}^2}{p_i^+} \,,
\end{eqnarray}
and
\begin{eqnarray}
	&&{d^3\tilde{p}_i}\equiv \frac{dp_i^+d^2p_{i\perp}}{2(2\pi)^3}\,, ~ \delta^3(\tilde{p})=\delta(p^+)\delta^2(p_{\perp})\,,
	 \nonumber\\
	&&|q_\alpha^a(\tilde{p},\lambda)\rangle=d^{\dagger a}_{\alpha}(\tilde{p},\lambda)|0\rangle\,,
	 ~\{d_{\alpha'}^{a'}(\tilde{p'},\lambda'),d^{\dagger a}_{\alpha}(\tilde{p},\lambda)\}=2(2\pi)^{3}\delta^3(\tilde{p'}-\tilde{p})\delta_{\lambda'\lambda}\delta_{\alpha'\alpha}\delta^{a'a}\,,
	\label{state}
\end{eqnarray}
respectively.
The internal motions of the constituent quarks are described by the  kinematic variables of $(q_{\perp},\xi)$,
 $(Q_{\perp},\eta)$ and $P_{tot}$, given by
\begin{eqnarray}
P_{tot}&=&\tilde{P}_1+\tilde{P}_2+\tilde{P}_3, \qquad \xi=\frac{p_1^+}{p_1^++p_2^+}, \qquad \eta=\frac{p_1^++p_2^+}{P_{tot}^+}\,,
 \nonumber \\
q_{\perp}&=&(1-\xi)p_{1\perp}-\xi p_{2\perp},\quad  Q_{\perp}=(1-\eta)(p_{1\perp}+p_{2\perp})-\eta p_{3\perp} \,,
\label{Lkin}
\end{eqnarray}
where $(q_{\perp},\xi)$ characterize the relative motion between the first and second quarks, while $(Q_{\perp},\eta)$  the third and other two quarks. 
The invariant masses of  $(q_{\perp},\xi)$ and $(Q_{\perp},\eta)$ systems are represented by~\cite{Schlumpf:1992ce}
\begin{eqnarray}
M_3^2=\frac{q_\perp^2}{\xi(1-\xi)}+\frac{m_1^2}{\xi}+\frac{m_2^2}{1-\xi}\,, \nonumber\\
M^2=\frac{Q_\perp^2}{\eta(1-\eta)}+\frac{M_3^2}{\eta}+\frac{m_3^2}{1-\eta}\,,
\end{eqnarray}
respectively.
The vertex function of $\Psi^{SS_z}(\tilde{p}_1,\tilde{p}_2,\tilde{p}_3,\lambda_1,\lambda_2,\lambda_3)$ in Eq.~(\ref{eq1}) 
can be written as~\cite{Lorce:2011dv,Zhang:1994ti,Schlumpf:1992ce}
\begin{eqnarray}
\Psi^{SS_z}(\tilde{p}_1,\tilde{p}_2,\tilde{p}_3,\lambda_1,\lambda_2,\lambda_3)&=&\Phi(q_\perp,\xi,Q_\perp,\eta)\Xi^{SS_z}(\lambda_1,\lambda_2,\lambda_3)\,,
\end{eqnarray}
where $\Phi(q_\perp,\xi,Q_\perp,\eta)$ is
 the momentum distribution of the constituent quarks  and $\Xi^{SS_z}(\lambda_1,\lambda_2,\lambda_3)$ represents
 the momentum-dependent  spin wave function, given by 
 \begin{eqnarray}
\Xi^{SS_z}(\lambda_1,\lambda_2,\lambda_3)&=&\sum_{s_1,s_2,s_3}\langle\lambda_1|R^{\dagger}_1|s_1\rangle\langle\lambda_2|R^{\dagger}_2|s_2\rangle\langle\lambda_3|R^{\dagger}_3|s_3\rangle \bigg\langle\frac{1}{2}s_1,\frac{1}{2}s_2,\frac{1}{2}s_3\bigg|SS_z\bigg\rangle\,,
\end{eqnarray}
with
 $\big\langle\frac{1}{2}s_1,\frac{1}{2}s_2,\frac{1}{2}s_3\big|SS_z\big\rangle$  the usual $SU(2)$ Clebsch-Gordan coefficient, and  $R_i$ the  Melosh transformation,   corresponding to the $i$th constituent quark, expressed by~\cite{Polyzou:2012ut,Schlumpf:1992ce}
\begin{eqnarray}
&&R_1=R_M(\eta,Q_\perp,M_3,M)R_M(\xi,q_\perp,m_1,M_3)\,, \nonumber\\
&&R_2=R_M(\eta,Q_\perp,M_3,M)R_M(1-\xi,-q_\perp,m_2,M_3) \,,\nonumber\\
&&R_3=R_M(1-\eta,-Q_\perp,m_3,M)\,,
\end{eqnarray}
with 
\begin{eqnarray}
R_M(x,p_{\perp},m,M)&=&\frac{m+xM-i\vec{\sigma}\cdot(\vec{n}\times \vec{p})}{\sqrt{(m+xM)^2+p_\perp^2}},
\end{eqnarray}
where $\vec{\sigma}$ stands for the Pauli matrix and $\vec{n}=(0,0,1)$.

In principle, one could solve the momentum wave function by introducing the QCD-inspired effective potential like the one-gluon exchange one.  
However, the equation will become too complicated to be solved in the three-body case. Therefore, we use the phenomenological Gaussian type wave function with some suitable shape parameters to describe the momentum distributions of the constituent quarks. 
It is also possible to compare the LFQM wave function with the light-cone distribution amplitude to get the QCD renormalization improvement at
 different energy scales~\cite{Bell:2013tfa}.
The baryon spin-flavor-momentum wave function $F_{abc}\Psi^{SS_z}(\tilde{p}_1,\tilde{p}_2,\tilde{p}_3,\lambda_1,\lambda_2,\lambda_3)$ should be totally symmetric under any permutations of the quarks to keep the Fermi statistics. The spin-flavor-momentum wave functions of $\Lambda_b^0$ and $p$ are given by
\begin{eqnarray}
	&&|\Lambda_b^0\rangle=\frac{1}{\sqrt{6}}[\phi_3\chi^{\rho3}(|dub\rangle-|udb\rangle)+\phi_2\chi^{\rho2}(|dbu\rangle-|ubd\rangle)+\phi_1\chi^{\rho 1}(|bdu\rangle-|bud\rangle)]\,,
	\nonumber\\
		&&|p\rangle=\frac{1}{3}\phi[\chi^{\rho3}(|udu\rangle-|duu\rangle)+\chi^{\rho2}(|uud\rangle-|duu\rangle)+\chi^{\rho1}(|uud\rangle-|udu\rangle)]\,,
		\label{sf}
\end{eqnarray}		
 where
 the explicit forms for  $\phi_{1,2}$ can be found   in  Ref.~\cite{Geng:2020fng},
\begin{eqnarray}
\phi_3={\cal N}\sqrt{\frac{\partial q_z}{\partial \xi}\frac{\partial Q_z}{\partial \eta}}e^{-\frac{{\bf Q}^2}{2\beta_{Q}^2}-\frac{{\bf q}^2}{2\beta_q^2}}\,,
		 \quad 		
		\chi^{\rho 3}_{\uparrow}=\frac{1}{\sqrt{2}}(|\uparrow\downarrow\uparrow\rangle-|\downarrow\uparrow\uparrow\rangle)\,,
\end{eqnarray}
and
\begin{eqnarray}
&& q_z=\frac{\xi M_3}{2}-\frac{m_1^2+q_\perp^2}{2M_3\xi}\,, \quad Q_z=\frac{\eta M}{2}-\frac{M_3^2+Q_\perp^2}{2M\eta} \nonumber\\
&& {\bf q}=(q_{\perp},q_z)\quad  {\bf Q}=(Q_{\perp},Q_z) \,,
\label{3kin}
\end{eqnarray}
with ${\cal N}=2(2\pi)^3(\beta_q\beta_{Q}\pi)^{-3/2}$ and  $\beta_{q,Q}$ being the normalized constant
and   shape parameters, respectively.

 Here, the baryon state is normalized as
\begin{eqnarray}
\langle {\bf B}&,P',S',S'_z|{\bf B}&,P,S,S_z\rangle=2(2\pi)^3P^+\delta^3(\tilde{P'}-\tilde{P})\delta_{S_z'S_z}\,,
\label{baryonN}
\end{eqnarray}
resulting in
the normalization of the momentum wave function, given by
\begin{eqnarray}
\frac{1}{2^2(2\pi)^6}	\int d\xi d\eta d^2q_{\perp}d^2Q_{\perp} |\phi_{3}|^2=1\,.
\end{eqnarray} 
We emphasize that the momentum wave functions of $\phi_i$ are associated with the different shape parameters of $\beta_q$ and $\beta_{Q}$ in $\Lambda_{b}^0$.  
 For the proton, the momentum distribution functions are the same, $i.e.$
 $\phi=\phi_3$ ($\beta_q=\beta_{Q}$), for any spin-flavor state because of the  isospin symmetry. 
 
\subsection{Baryonic Transition form factors}
The baryonic transition form factors  of the $V-A$  weak current  are given by Eq.~(\ref{form})
with $P'-P=k$.
We choose the frame such that $P^+$ is conserved ($k^+=0,k^2=-k_\perp^2$) to calculate the form factors  in order to avoid  other diagrams involving particle productions, also known as the Z-graphs in the light-front formalism~\cite{Schlumpf:1992ce,Zhang:1994ti}.
The matrix elements of the vector and axial-vector currents at quark level correspond to three different effective diagrams 
as shown in Fig.~1. 
\begin{figure}
 \begin{minipage}[h]{0.3\linewidth}
 	 	(a)
 	\centering
 	\includegraphics[width=2in]{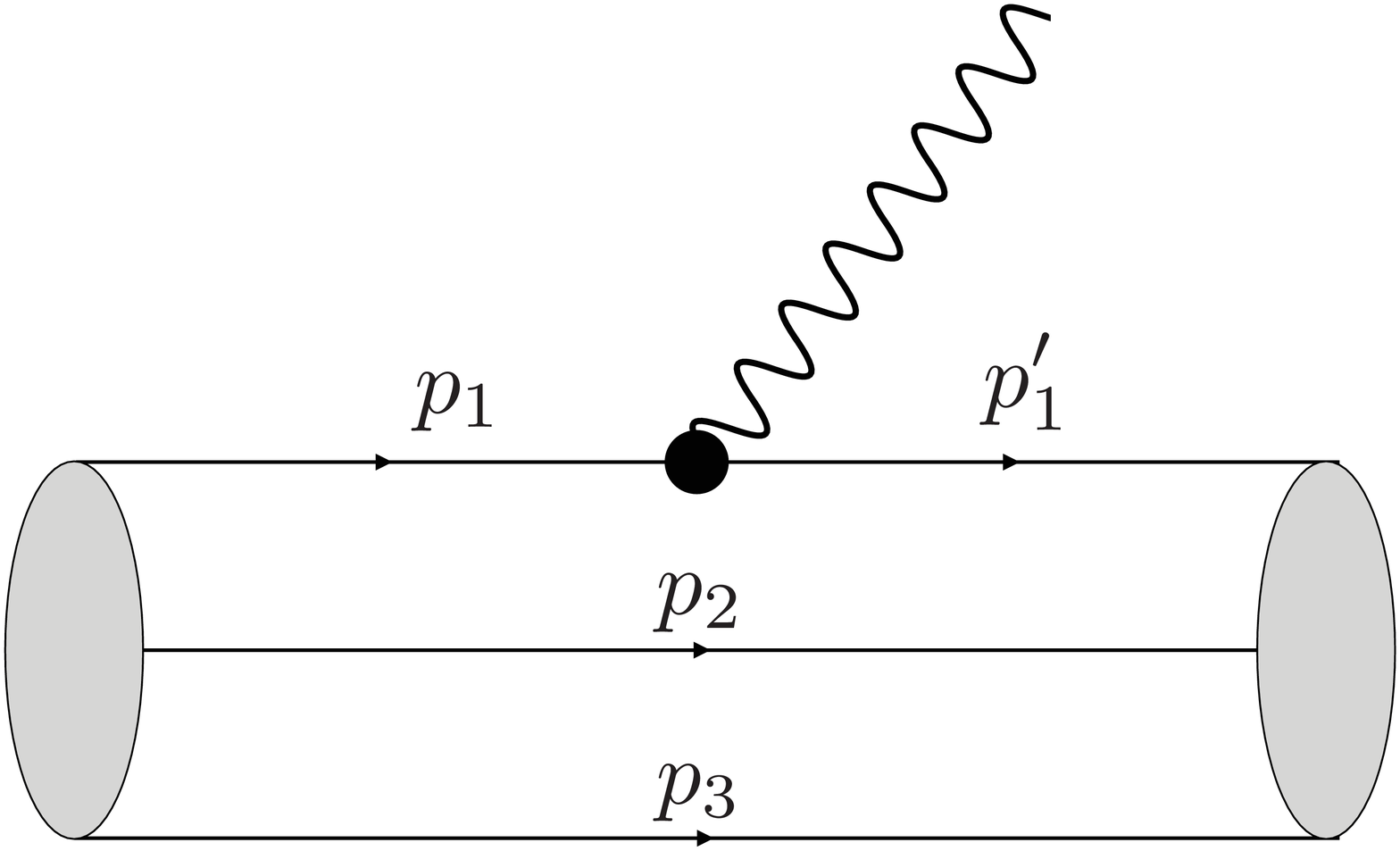}
 \end{minipage}
\begin{minipage}[h]{0.3\linewidth}
	(b)
 	\centering
\includegraphics[width=2in]{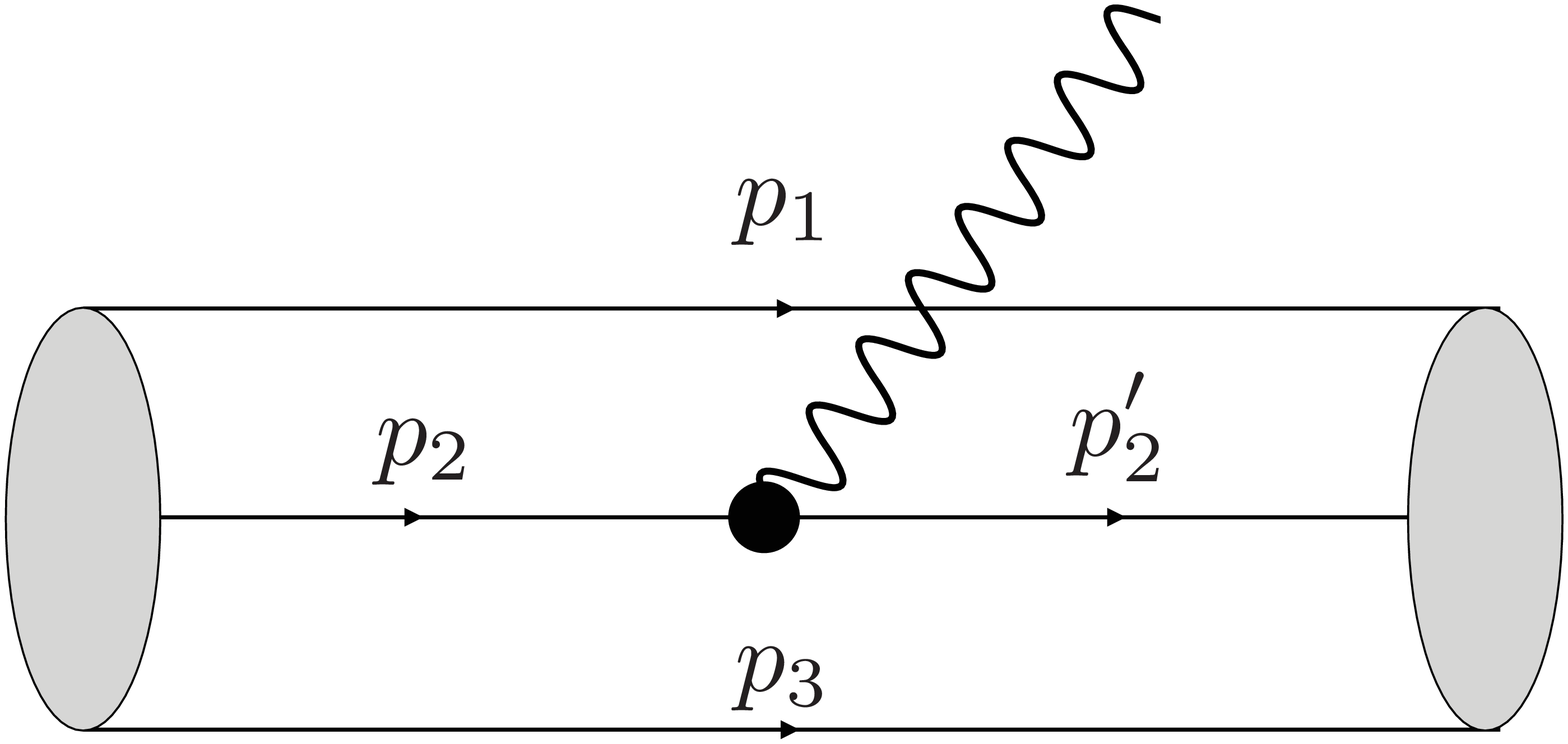}.
\end{minipage}
\begin{minipage}[h]{0.3\linewidth}
	(c)
 	\centering
\includegraphics[width=2in]{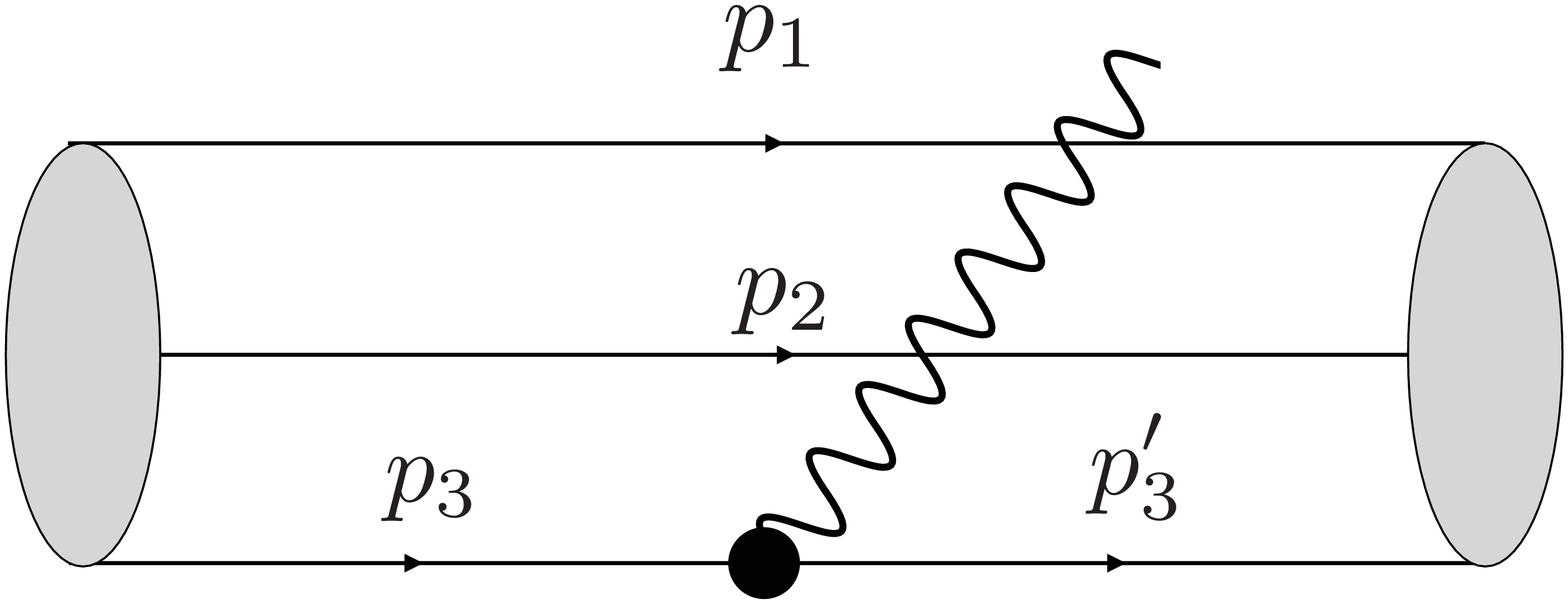}
\end{minipage}
\caption{Feynman diagrams for the baryonic weak transitions, where the bullet of ``$\bullet$'' denotes the V-A  current vertex,
with  (a) $p'_1-p_1=k$, (b) $p'_2-p_2=k$ and (c) $p'_3-p_3=k$.}
\end{figure}
Since the spin-flavor-momentum wave functions of the baryons are totally symmetric under the permutation of  the quarks indices, 
we have that $(a)+(b)+(c)=3(a)=3(b)=3(c)$~\cite{Schlumpf:1992ce}. 
We only present the calculation for  the diagram (c) in Fig.~1, 
which contains simpler and cleaner forms with our notation $(q_{\perp},Q_{\perp},\xi,\eta)$.
We can extract the form factors from the matrix elements through the relations,
\begin{eqnarray}
	f_1(k^2)&=&\frac{1}{2P^+}\langle {\bf B}_f,P',\uparrow|\bar{q}\gamma^{+}c|{\bf B}_i,P,\uparrow\rangle\,, \nonumber \\
	f_2(k^2)&=&\frac{1}{2P^+}\frac{M_{{\bf B}_i}}{k_{\perp}}\langle {\bf B}_f,P',\uparrow|\bar{q}\gamma^{+}c|{\bf B}_i,P,\downarrow\rangle\,, \nonumber \\
	g_1(k^2)&=&\frac{1}{2P^+}\langle {\bf B}_f,P',\uparrow|\bar{q}\gamma^{+}\gamma_5c|{\bf B}_i,P,\uparrow\rangle\,, \nonumber \\
	g_2(k^2)&=&\frac{1}{2P^+}\frac{M_{{\bf B}_i}}{k_{\perp}}\langle {\bf B}_f,P',\uparrow|\bar{q}\gamma^{+}\gamma_5|{\bf B}_i,P,\downarrow\rangle  \,.
	\label{fm}
\end{eqnarray}
Note that $f_3$ and $g_3$ are unobtainable when $k^+=0$, but  they are irrelevant in this work
because of the kinematic suppression of $m_P^2/m_{\Lambda_b}^2$ associated with them.
The full expressions of the form factors in Eq.~(\ref{fm}) are give by 
\begin{eqnarray}
\label{f1}
&&f_1(k^2)=\frac{3}{2^2(2\pi)^6}\int d\xi d\eta d^2q_\perp d^2Q_\perp\Phi(q'_{\perp},\xi,Q'_\perp,\eta)\Phi(q_{\perp},\xi,Q_\perp,\eta)(F^{def} F_{abc}\delta_{q_{f}q}\delta_{cq^c}\delta_{d}^{a}\delta_{e}^{b})\nonumber\\
&&\times \sum_{s_1,s_2,s_3}\sum_{s'_1,s'_2,s'_3}\langle S',\uparrow|s'_1,s'_2,s'_3\rangle\langle s_1,s_2,s_3|S,\uparrow\rangle
\prod_{i=1,2,3}\langle s'_i|R'_iR^{\dagger}_i|s_i\rangle \,,
\end{eqnarray}
\begin{eqnarray}
\label{g1}
&&g_1(k^2)=\frac{3}{2^2(2\pi)^6}\int d\xi d\eta d^2q_\perp d^2Q_\perp\Phi(q'_{\perp},\xi,Q'_\perp,\eta)\Phi(q_{\perp},\xi,Q_\perp,\eta)(F^{def} F_{abc}\delta_{q_{f}q}\delta_{cq^c}\delta_{d}^{a}\delta_{e}^{b})\nonumber\\
&&\times \sum_{s_1,s_2,s_3}\sum_{s'_1,s'_2,s'_3}\langle S',\uparrow|s'_1,s'_2,s'_3\rangle\langle s_1,s_2,s_3|S,\uparrow\rangle
\prod_{i=1,2}\langle s'_i|R'_iR^{\dagger}_i|s_i\rangle \langle s'_3|R'_3\sigma_zR^{\dagger}_3|s_3\rangle\,,
\end{eqnarray}
\begin{eqnarray}
\label{f2}
&&f_2(k^2)=\frac{3}{2^2(2\pi)^6}\frac{M_{{\bf B}_i}}{k_{\perp}}\int d\xi d\eta d^2q_\perp d^2Q_\perp\Phi(q'_{\perp},\xi,Q'_\perp,\eta)\Phi(q_{\perp},\xi,Q_\perp,\eta)(F^{def} F_{abc}\delta_{q_{f}q}\delta_{cq^c}\delta_{d}^{a}\delta_{e}^{b})\nonumber\\
&&\times \sum_{s_1,s_2,s_3}\sum_{s'_1,s'_2,s'_3}\langle S',\uparrow|s'_1,s'_2,s'_3\rangle\langle s_1,s_2,s_3|S,\downarrow\rangle
\prod_{i=1,2,3}\langle s'_i|R'_iR^{\dagger}_i|s_i\rangle  \,,
\end{eqnarray}
\begin{eqnarray}
&&g_2(k^2)=\frac{3}{2^2(2\pi)^6}\frac{M_{{\bf B}_i}}{k_{\perp}}\int d\xi d\eta d^2q_\perp d^2Q_\perp\Phi(q'_{\perp},\xi,Q'_\perp,\eta)\Phi(q_{\perp},\xi,Q_\perp,\eta)(F^{def} F_{abc}\delta_{q_{f}q}\delta_{cq^c}\delta_{d}^{a}\delta_{e}^{b})\nonumber\\
&&\times \sum_{s_1,s_2,s_3}\sum_{s'_1,s'_2,s'_3}\langle S',\uparrow|s'_1,s'_2,s'_3\rangle\langle s_1,s_2,s_3|S,\downarrow\rangle
\prod_{i=1,2}\langle s'_i|R'_iR^{\dagger}_i|s_i\rangle \langle s'_3|R'_3\sigma_zR^{\dagger}_3|s_3\rangle\,.
\label{g2}  
\end{eqnarray}

\section{Numerical Results}
We choose   the Wolfenstein parameterization for the CKM matrix with the corresponding parameters, taken to be~\cite{pdg}
\begin{eqnarray}
	&\lambda=0.22650\pm0.00048,\quad A=0.790^{+0.017}_{-0.012}, \quad \rho=0.141^{+0.016}_{-0.017},\quad \eta=0.357 \pm 0.011 \,.
\end{eqnarray}
Under the naive dimensional regularization (NDR) scheme and $\overline{\text{MS}}$ renormalization, the  numerical values
of the effective Wilson coefficients  $c_i^{eff}(\mu)$ at  $\mu=2.5\text{ GeV}$ for both $b\to q$ and $\bar{b}\to \bar{q}$ 
are given in Table.~\ref{wilson}~\cite{Ali:1998eb,Buras:1991jm}.  
\begin{table}[t]
	\caption{Numerical values of the effective Wilson coefficients with the NDR scheme and $\overline{\text{MS}}$ at the renormalization scale of
	$\mu=2.5\text{ GeV}$}
	\begin{tabular}{lcc|cc}
		\hline \hline
		&$b\to d$& $\bar{b}\to \bar{d}$&$b\to s$&$\bar{b}\to \bar{s}$\\
		\hline
		$c_1^{eff}$&$1.168$&$1.168$&$1.168$&$1.168$\\
		$c_2^{eff}$&$-0.365$&$-0.365$&$-0.365$&$-0.365$\\
		$10^4 c_3^{eff}$&$238+14i$&$254+43i$&$243+31i$&$241+32i$\\
		$10^4 c_4^{eff}$&$-497-42i$&$-545-130i$&$-512-94i$&$-506-97i$\\
		$10^4 c_5^{eff}$&$145+14i$&$162+43i$&$150+31i$&$148+32i$\\
		$10^4 c_6^{eff}$&$-633-42i$&$-682-130i$&$-649-94i$&$-643-97i$\\
		$10^4 c_7^{eff}$&$-1.0-1.0i$&$-1.4-1.8i$&$-1.1-2.2i$&$-1.1-1.3i$\\
		$10^4 c_8^{eff}$&$5.0$&$5.0$&$5.0$&$5.0$\\
		$10^4 c_9^{eff}$&$-112-1.3i$&$-113-2.7i$&$-112-2.2i$&$-112-2.2i$\\
		$10^4 c_{10}^{eff}$&$20$&$20$&$20$&$20$\\
		\hline
		\hline
	\end{tabular}
	\label{wilson}
\end{table}
For the quark masses  in Eq.~(\ref{chiral enhancement}), we start from the values at $\mu=2.0\text{ GeV}$ given by PDG~\cite{pdg}, 
and use the renormalization group equation~\cite{Buras:1991jm} to determine them at  $\mu=2.5\text{ GeV}$, given by
\begin{eqnarray}
	m_b(2.5\text{ GeV})=4.88\text{ GeV}\quad m_s(2.5\text{ GeV})=88.1\text{ MeV} \nonumber\\
	m_d(2.5\text{ GeV})=4.42\text{ MeV}\quad m_u(2.5\text{ GeV})=2.04\text{ MeV} 
\end{eqnarray}
In Table.~\ref{par}, we show our parameter sets of the hadron masses and meson decay constants.
\begin{table}
	\caption{Hadron masses and meson decay constants in  units of GeV}
	\begin{tabular}{cccccccccc}
		\hline
		\hline
	$m_{\Lambda_{b}}$&$m_{p}$&$m_{\pi}$&$m_{K}$&$m_{\rho}$& $m_{K^*}$&$f_{\pi}$&$f_{K}$&$f_{\rho}$&$f_{K^*}$\\
		\hline
		$5.62$&$0.94$&$0.14$&$0.49$&$0.77$&$0.89$&$0.13$&$0.16$&$0.21$&$0.22$\\
		\hline
		\hline
	\end{tabular}\label{par}
\end{table}
In our numerical calculations, we also set the $\Lambda_{b}^0$ life time to be $\tau_{\Lambda_b^0}=146.4\text{ fs}$~\cite{pdg}. 
In this work,  the effective color number 
$N_c^{eff}$ is fixed to be 3.

Since the baryonic transition form factors in LFQM can be only evaluated  in the space-like region ($k^2=-k^2_\perp$) due to the condition $k^+=0$,
 we take some analytic functions to fit $f_{1(2)}(k^2)$ and $g_{1(2)}(k^2)$ in the space-like region and
perform  their analytical continuations to the physical time-like region $(k^2>0)$ as in Refs.~\cite{Jaus:1991cy,Geng:2020fng,Jaus:1996np,Cheng:2003sm,Ke:2007tg,Cheng:2004cc}. 
We employ the numerical values of the constituent quark masses and shape parameters in Table.~\ref{sh}. 
 
 \begin{table}
\caption{ Model parameters in LFQM with the constituent quark masses ($m_i$) and shape parameter ($\beta_{q{\bf B}}$ and $\beta_{{Q\bf B}}$) in
 units of GeV.} 
\begin{tabular}{cccccc}
\hline
\hline
$m_b$&$m_d$&$m_u$&$\beta_{Q\Lambda_b}$&$\beta_{q\Lambda_b}$&$\beta_{qp}$\\
\hline
4.88&0.26&0.26&0.601&0.365&0.365\\
\hline
\hline
\end{tabular}
\label{sh}
\end{table} 
  By using  Eqs.~(\ref{f1})-(\ref{g2}), we compute totally 32 points for all form factors from $k^2=0$ to $k^2=-9.7\text{ GeV}^2$.
To fit the $k^2$ dependences of the form factors in the space-like region,
we use the form, given by
\begin{eqnarray}
	F(k^2)=\frac{F(0)}{1-q_1k^2+q_2k^4} \,,
\end{eqnarray}
where $q_{1,2}$ are the fitting parameters.
Our results for the form factors are presented  in Table.~\ref{nf} and compared with the results from light-cone sum rule (LCSR)~\cite{Khodjamirian:2011jp} where the form factors are parameterized by
\begin{eqnarray}
	f_{i}\left(k^{2}\right)&=&\frac{f_{i}(0)}{1-q^{2} / m_{B^{*}\left(1^{-}\right)}^{2}}\bigg\{1+b_{i}\left(z\left(k^{2}, t_{0}\right)-z\left(0, t_{0}\right)\right)\bigg\}\,, \nonumber\\
	g_{i}\left(k^{2}\right)&=&\frac{g_{i}(0)}{1-q^{2} / m_{B^{*}\left(1^{+}\right)}^{2}}\bigg\{1+\tilde{b}_{i}\left(z\left(k^{2}, t_{0}\right)-z\left(0, t_{0}\right)\right)\bigg\}
\end{eqnarray}
The definition of these parameters can be found in Ref.~\cite{Khodjamirian:2011jp}.
\begin{table}
	\caption{Form factors for the transition matrix elements of $\Lambda_b^0 \to p$ in LFQM as well as LCSR with the axial-vector
	  (pseudo-scalar) interpolating current of  $\eta_{\Lambda_{b}}^{{\cal A}({\cal P})}$~\cite{Khodjamirian:2011jp}.}
\begin{tabular}{ccccc}
	\hline
	\hline
	&$f_1$&$f_2$&$g_1$&$g_2$\\
		\hline
	$F(0)$&$0.132\pm0.003$&$0.149\pm0.003$&$0.129\pm0.003$&$-0.020\pm0.001$\\
	$q_1$&$0.802\pm0.199$&$0.708\pm0.189$&$0.7702\pm0.187$&$0.036\pm0.137$\\
	$q_2$&$1.164\pm0.208$&$1.082\pm0.109$&$1.103\pm0.190$&$0.732\pm0.107$\\
	\hline
		\hline
		&\multicolumn{2}{c}{$\eta_{\Lambda_{b}}^{(\cal A)}$}&\multicolumn{2}{c}{$\eta_{\Lambda_{b}}^{(\cal P)}$}\\
	\hline
	$f_1(0)$&\multicolumn{2}{c}{$0.14^{+0.03}_{-0.03}$}&\multicolumn{2}{c}{$0.12^{+0.03}_{-0.04}$}\\
	$b_1$&\multicolumn{2}{c}{$-1.49^{+1.68}_{-1.88}$}&\multicolumn{2}{c}{$-9.13^{+0.88}_{-1.12}$}\\
	$f_2(0)$\footnote[1]{The authors in Ref.~\cite{Khodjamirian:2011jp} use  the opposite sign convention for  $f_2(g_2)$. }&\multicolumn{2}{c}{$-0.054^{+0.016}_{-0.013}$}&\multicolumn{2}{c}{$-0.047^{+0.015}_{-0.013}$}\\
		$b_2$&\multicolumn{2}{c}{$-14.0^{+1.2}_{-1.8}$}&\multicolumn{2}{c}{$-18.5^{+1.7}_{-2.0}$}\\
	$g_1(0)$&\multicolumn{2}{c}{$0.14^{+0.03}_{-0.03}$}&\multicolumn{2}{c}{$0.12^{+0.03}_{-0.03}$}\\
		$\tilde{b}_1$&\multicolumn{2}{c}{$-4.05^{+1.38}_{-1.81}$}&\multicolumn{2}{c}{$-9.18^{+0.75}_{-1.06}$}\\
	$g_2(0)^{\text{a}}$&\multicolumn{2}{c}{$-0.028^{+0.012}_{-0.009}$}&\multicolumn{2}{c}{$-0.016^{+0.007}_{-0.005}$}\\
		$\tilde{b}_2$&\multicolumn{2}{c}{$-20.2^{+1.0}_{-2.1}$}&\multicolumn{2}{c}{$-22.5^{+1.3}_{-1.7}$}\\
		\hline
	\hline
\end{tabular}
\label{nf}
\end{table}
From the table, one has that $f_1\simeq g_1$,  which agrees with the relation of  $f_1= g_1$ in the heavy quark limit. 
One can also see that   $f_2(k^2=0)>f_1(k^2=0)$, which is  similar to the cases  in the $\Lambda_c^+$  decays~\cite{Geng:2020fng}, but
different from  those in LCSR~\cite{Khodjamirian:2011jp}  for the $\Lambda_b^0$ transitions. 
For  LFQM and  LCSR,
 the results of $f_1(g_1)$   are the same, whereas those of $f_2(g_2)$  are different, in  the low-momentum transfer region. 
 The predictions of the branching ratios and CP-asymmetries in these two models are similar because the contributions of $f_1$ and $g_1$ are dominated 
 in $\Lambda_{b}^0 \to p P(V)$. The discrepancies between LFQM and LCSR  appear in the $\Lambda_{b}^0$ semi-leptonic decays. 
 The decay rates in these semi-leptonic decays for LFQM 
 are much smaller than the LCSR ones as well as the data due to the lack of the
information from the $B^{*}(1^{-(+)})$ poles in LFQM, which are important in the calculations of $\Lambda_{b}^0\to p \ell^-\bar{\nu_{\ell}}$ via the analytical continuation method.
Our predictions of the decay branching ratios and direct CP-violating rate asymmetries are summarized in Table~\ref{com},
where the first and second errors contain the uncertainties from the form factors and  Wolfenstein parameters in Table~\ref{nf}, respectively.  It is interesting to see that
our results of $\mathcal{A}_{CP}(\Lambda_{b}^0\to p V)$ are free of the baryonic uncertainties under the generalized factorization assumption.
The reason is that all form-factor dependencies in $\Gamma(\Lambda_{b}^0\to p V)$ can be totally factorized out within the generalized factorization framework so that they get canceled for the CP-violating rate asymmetries of  $\mathcal{A}_{CP}(\Lambda_{b}^0\to p V)$~\cite{HG,Hsiao:2017tif}. As a result, the numerical values of $\mathcal{A}_{CP}(\Lambda_{b}^0\to p V)$ are QCD model-independent. 
In the table, we also show 
 the recent experimental data~\cite{pdg} as well as  other theoretical calculations in the literature~\cite{Hsiao:2017tif,pQCD,Zhu:2018jet,Zhao:2018zcb},
where
 $\text{pQCD}_a$ and $\text{pQCD}_b$  represent  the conventional and  hybrid pQCD approaches~\cite{pQCD}, 
 while
   $\text{LFQM}_a$~\cite{Zhu:2018jet} and $\text{LFQM}_b$~\cite{Zhao:2018zcb} refer to those in LFQM
   with and without the considerations of  the QCD and electroweak-penguin loop contributions, respectively. 
\begin{table}
	\caption{ Our results in comparison with the experimental data and those in various theoretical calculations in the literature,
	where  $\mathcal{B}$s and $\mathcal{A}$s are in units of $10^{-6}$ and $\%$, respectively. In our results, the first errors come from hadronic uncertainties and the second errors come from the uncertainties of CKM matrix. 
	}
	\label{com}
	{	\footnotesize
		\begin{tabular}{c|cccc}
		\hline\hline
			&$\mathcal{B}(\Lambda_{b}^0\to p \pi^- )$ &$\mathcal{B}(\Lambda_{b}^0\to p K^- )$ &$\mathcal{B}(\Lambda_{b}^0\to p \rho^- )$ &$\mathcal{B}(\Lambda_{b}^0\to p K^{*-})$ \\
			\hline
			this work&$4.18\pm0.15\pm0.30$&$5.76\pm0.88\pm0.23$&$12.13\pm3.27\pm0.91$&$2.58\pm0.87\pm0.13$\\
			Data~\cite{pdg}&$4.5\pm0.8$& $5.4\pm1.0$&-&-\\
			$\text{GFA}$~\cite{Hsiao:2017tif}&$4.25^{+1.04}_{-0.48}\pm0.93$&$4.49^{+0.84}_{-0.39}\pm0.64$&$11.03^{+2.72}_{-1.25}\pm2.45$&$2.86^{+0.62}_{-0.29}\pm0.52$\\
			$\text{pQCD}_a$~\cite{pQCD}&4.66&1.82&-&-\\
			$\text{pQCD}_b$~\cite{pQCD}&5.21&2.02&-&-\\
			$\text{LFQM}_a$~\cite{Zhu:2018jet}&4.30&2.17&7.47&1.01\\
			$\text{LFQM}_b$~\cite{Zhao:2018zcb}&8.90&0.718&26.1&1.34\\
			MBM~\cite{Geng:2020ofy}& 4.5&3.4&-&-\\
			LCSR~\cite{Khodjamirian:2011jp}&$3.8^{+1.3}_{-1.0}(2.8^{+1.1}_{-0.9})$&-&-&-\\
		\hline\hline	
		&$\mathcal{A}_{CP}(\Lambda_{b}^0\to p \pi^- )$ &$\mathcal{A}_{CP}(\Lambda_{b}^0\to p K^- )$ &$\mathcal{A}_{CP}(\Lambda_{b}^0\to p \rho^- )$ &$\mathcal{A}_{CP}(\Lambda_{b}^0\to p K^{*-})$ \\
		\hline
		this work&$-3.60\pm0.14\pm0.14$&$6.36\pm0.21\pm0.18$&$-3.32\pm0.00\pm0.14$&$19.25\pm0.00\pm0.80$\\
		Data~\cite{pdg}&$-2.5\pm2.9$& $-2.5\pm2.2$&-&-\\
		$\text{GFA}$~\cite{Hsiao:2017tif}&$-3.9^{+0.0}_{-0.0}\pm0.4$&$6.7^{+0.3}_{-0.2}\pm0.3$&$-3.8^{+0.0}_{-0.0}\pm0.4$&$19.7^{+0.4}_{-0.3}\pm1.4$\\
		$\text{pQCD}_a$~\cite{pQCD}&-32&-3&-&-\\
		$\text{pQCD}_b$~\cite{pQCD}&-31&-5&-&-\\
		$\text{LFQM}_a$~\cite{Zhu:2018jet}&$-3.37^{+0.29}_{-0.37}$&$10.1^{+1.3}_{-2.0}$&$-3.19^{+0.25}_{-0.25}$&$31.1^{+2.8}_{-1.9}$\\
		$\text{LFQM}_b$~\cite{Zhao:2018zcb}&-&-&-&-\\
		MBM~\cite{Geng:2020ofy}& -4.4&6.7&-&-\\
		\hline\hline	
		\end{tabular}
	}
\end{table}
We find that our predictions of ${\cal B}( \Lambda_b^0 \to p \pi^-,pK^-)$ are consistent with the experimental data and those in GFA and LCSR.
In addition, we have that ${\cal B}( \Lambda_b^0 \to p K^-) >{\cal B}( \Lambda_b^0 \to p \pi^-)$ as indicated by the data, which
is different from those predicted by pQCD~\cite{pQCD}, LFQM~\cite{Zhu:2018jet,Zhao:2018zcb}  and MBM~\cite{Geng:2020ofy}.
Our results for ${\cal B}( \Lambda_b^0 \to (p \rho^-,pK^{*-}))$ also agree with those in GFA~\cite{Hsiao:2017tif}.
 Note that $\text{LFQM}_b$~\cite{Zhao:2018zcb} does not consider the loop contributions, so that its results are incompatible with the data as well as  all  other theoretical ones.
 For the CP-violating rate asymmetries, we find that $\mathcal{A}_{CP}(\Lambda_{b}^0\to p \pi^- )=(-3.60\pm0.14\pm0.14)\%$, consistent with the data and 
 those from GFA~\cite{Hsiao:2017tif}, $\text{LFQM}_a$~\cite{Zhu:2018jet} and MBM~\cite{Geng:2020ofy}, but different from pQCD~\cite{pQCD}.
On the other hand, we obtain a positive  CP-violating rate asymmetry for  $ \Lambda_b^0 \to pK^-$, which is the same as GFA, $\text{LFQM}_a$ and MBM,
whereas the data of $(-2.5\pm2.2)\%$~\cite{pdg} along with those from pQCD is negative.
It is interesting to note that, similar to GFA and $\text{LFQM}_a$, our result in LFQM also predicts a  sizable  asymmetry of $\sim 20\%$ in $\Lambda_{b}^0\to p K^{*-}$.

\section{Conclusions}
We have studied the non-leptonic two body decays of $\Lambda_b^0 \to p M$ with LFQM based on the generalized factorization ansatz.  
By considering the Fermi statistic between quarks and determining spin-flavor structures  in baryons, we have evaluated the baryonic form factors
in the LFQM. In particular, we have found that  $f_1\simeq g_1$, which  agrees with the requirement  in the heavy quark limit,
whereas  $f_2(k^2=0)>f_1(k^2=0)$, 
 different from those in the  literature for the $\Lambda_b^0$ transitions. 
 It is possible to include the QCD renormalization effect by comparing the LFQM wave function and light-cone distribution amplitude to understand the uncertainty at different energy scales as the study in Ref.~\cite{Bell:2013tfa},
which  could be done in our future work. 
 	We have compared our form factors with those in LCSR, and shown that they are almost the same in the low-momentum transfer region. However, the LCSR ones give  better results than those in LFQM in the high-momentum transfer region because the analytical continuation method in LFQM misses the information about the $B^{*}(1^{+(-)})$ poles, which is important in the calculation for $\Lambda_{b}^0\to p \ell^- \bar{\nu_{\ell}}$.
From these baryonic  form factors in LFQM, we have found that
${\cal B}( \Lambda_b^0 \to p \pi^- ,pK^-)=(4.18\pm0.15\pm0.30, 5.76\pm0.88\pm0.23)\times10^{-6}$ and
${\mathcal{A}_{CP}}( \Lambda_b^0 \to p \pi^- ,pK^-)=(-3.60\pm0.14\pm0.14, 6.36\pm 0.21\pm0.18)\%$ 
in comparison  with the data of ${\cal B}( \Lambda_b^0 \to p \pi^- ,pK^-)=(4.5\pm0.8, 5.4\pm1.0)\times10^{-6}$ and
${\mathcal{A}_{CP}}( \Lambda_b^0 \to p \pi^- ,pK^-)=(-2.5\pm 2.9, -2.5\pm2.2)\%$~\cite{pdg}, respectively,
 while 
${\cal B}(  \Lambda_b^0 \to p \rho^-,pK^{*-} )=(12.13\pm3.27\pm0.23, 2.58\pm0.87\pm0.13)\times 10^{-6}$ and
${\mathcal{A}_{CP}}( \Lambda_b^0 \to p \rho^-,pK^{*-} )=(-3.32\pm0.00\pm0.14,19.25\pm0.00\pm0.80)\%$,  which could be observed by the  experiments at  LHCb.


\section*{ACKNOWLEDGMENTS}
This work was supported in part by National Center for Theoretical Sciences and 
MoST (MoST-107-2119-M-007-013-MY3).

\end{document}